\documentclass[12pt, prd, reprint, amsmath, amssymb, aps, superscriptaddress]{revtex4-1}
\usepackage{appendix}
\usepackage{color}
\usepackage[dvipsnames]{xcolor}
\usepackage{graphicx}
\usepackage[citecolor=blue, linkcolor=blue, urlcolor=blue, colorlinks=true]{hyperref}

\def\beq{\begin{eqnarray}}
\def\eeq{\end{eqnarray}}
\def\beqa{\begin{eqnarray}}
\def\eeqa{\end{eqnarray}}

\definecolor{darkred}{rgb}{.743,0,0}

\begin{document}
\title{The academic career in physics as a ``deal":  Choosing physics within a gendered power structure and excellence as an extra hurdle for women}

\author{Meytal Eran-Jona$^{1a}$, and Yosef Nir$^{2b}$}
\affiliation{$^1$Feinberg Graduate School, Weizmann Institute of Science, Rehovot, Israel 7610001\\
$^2$Department of Particle Physics and Astrophysics, Weizmann Institute of Science, Rehovot, Israel 7610001}
\email{$^a$meytal.jona@weizmann.ac.il; $^b$yosef.nir@weizmann.ac.il}
%\date{\today}

\begin{abstract}
The absence of women among academic staff in physics is in the focus of our research. To explore the causes of this gender imbalance, we conducted a nationwide representative survey among Ph.D. students and interviews with Ph.D. students and postdoctoral fellows. Studying both context factors and agency, we reveal the multiple and hidden ways in which gender operates as a power structure, putting up barriers to women's academic careers.
This latent power structure influences women's decision-making and experiences in several ways. In the academic field, it produces unequal competition in a male dominated playground. In the social sphere, choosing a demanding academic career is seen as disrupting gender order. Within the family, women carry a greater burden of family work and give precedence to their partner's career and preferences. Within this social structure, women who decide to follow an academic career feel that they must excel. The demand for excellence acts as an invisible barrier within the gender power structure that prevents talented women from pursuing an academic career in physics.

\end{abstract}
%\pacs{}

\maketitle
%\tableofcontents

%%%%%%%%%%%%%%%%%%%
\section{Introduction}
The small proportion of women among university students and academic staff in math-intensive disciplines is an issue of much concern. In the last two decades, much research has been carried out in this regard, and much effort invested in trying to improve the situation. Of particular interest is the field of physics, where the gender imbalance in the academy is particularly severe, and where women participation has shown no significant increase in spite of dedicated efforts in the US \cite{porter2019} and in Europe \cite{cochran2019,blue2018}.

Our study focusses on gender imbalance in the field of physics in the academy in Israel. The situation in Israel in this regard is interesting in several aspects. First, as we explain below, there are some relevant unique characteristics of the Israeli society, which is very familial. Second, in Israel, the percentages of women among physics students and academic staff are even poorer than in the US and western Europe: 16\% among graduate students and 6\% among faculty members  \cite{jona2019}. Third, these small rates are even more striking when compared to other fields in the Israeli academy, such as medicine, where women constitute 69\% among graduate students and 35\% among faculty members, or biology, where the proportion is 58\% and 30\%, respectively.

Looking at these data, we asked ourselves why so few women pursue an academic career in physics. As a first step on the way to answer this question, we set to learn about the decision making of students, both male and female, when considering an academic career trajectory. Therefore we pose the following three research questions: How do Ph.D. students perceive the academic career path? What are the female student considerations in going for a post-doctoral training abroad? In what ways are women's considerations for and against pursuing an academic career shaped (constructed) by gender? To answer these questions we choose a mixed methods research methodology, combining a representative quantitative survey of all physics Ph.D. students in Israel, with qualitative tools, conducting interviews with female Ph.D. students and post doctoral fellows.

In this paper we present the stories of these young women, while examining their considerations and trying to untangle the impact of the context -- their personal, professional, economic and family circles -- on their decision making. Women's perspective is reviewed against the ``deal" that the academic career offers them.

%%%%%%%%%%%%%%%%%%%
\section{Intersecting knowledge fields}
In the following we review the literature regarding obstacles and difficulties that graduate students in physics face along their studies, with some references to other STEM fields and to research on academic staff. In particular, we focus on gender differences and on the challenges for female physicists within academy.

The research that deals with the integration of women into academic careers is embedded in various academic disciplines, which are mostly studied separately: psychology, sociology of education, sociology of organizations, gender studies, labor studies, economics and more. The starting point for our study lies in the understanding that the answers to complex questions about barriers to women's integration within academia cannot be found in a single theoretical field or discipline. We must step beyond these traditional disciplinary borders toward interdisciplinary research, which can provide a diverse and multidimensional perspective to examine and answer the research questions.

The lack of interdisciplinary research among those dealing with career issues and the need to integrate perspectives have been highlighted as early as the 1970s \cite{vanmannen1977}. The lack of interdisciplinary research in the field results in partial and biased understanding of careers, and a multilayered approach is needed \cite{chudzikowski2010}.

In response to this challenge, we study career decision making in physics as a multilayered and multi-dimensional phenomenon, combining in our analysis contextual, organizational and individual variables and their intersection. Looking at both structure and agency, we do not view careers as based solely on an individual's free choice, but rather on the intersection between the individual choice, the roles and structure of physics as a field (in terms of Bourdieu), the gendered social expectations and norms,  and the family division of role and employment expectations.

%%%%%%%%%%%%%
\subsection{Gender and power}
The understanding that power is an integral part of gender relations is central to the feminist theory \cite{connell1987,hartsock1987,komter1991,scott1986,sheppard1989}. Power is an integral part of Scott's definition of ``gender" \cite{scott1986}. According to Scott, ``gender is not only a constitutive element of social relationships based on perceived differences between the sexes, but it is also a primary way of signifying relationships of power. Changes in the organization of social relationships always occur concurrently with changes in representations of power."

The idea that social structure and processes are gendered was developed within the feminist discourse from early 1980s. Our theoretical starting point is based on the understanding that gender is a power structure in society, within family and in organizations \cite{scott1986,kanter1977,acker1990,connell1987,connell1990}.

In her book ``Gender and Power" \cite{connell1987}, Connell characterizes the elusive way in which power works in social relationships. ``While particular transactions involving power are easy to observe, it is often difficult to see beyond individual acts of force or oppression to structure of power, a set of social relations with some scope and permanence." This elusive social power may be ``a balance of advantage or an inequality of resources in a workplace, a household, or a larger institution". Connell uses the term ``gender regime" to analyze gender relations within the family \cite{connell1987,connell1990}.

Acker claims that gender is a structural feature of labor organizations \cite{acker1990}. She proposes a theory of organizations and gender that follows Kanter \cite{kanter1977} and other scholars \cite{connell1987,reskin1987,scott1986} that claim that gender is a structural feature of organizations and not a characteristic feature of individuals bringing it to the workplace. Acker shows how deeply gender is embedded within organizations. She claims that ``the organizational structure is not gender-neutral, on the contrary, assumptions about gender underlie the documents and contracts used to construct organizations and to provide the commonsense ground for theorizing them". According to Acker, when we say that an organization is gender biased, it means that advantage and disadvantage, exploitation and control, action and emotion, meaning and identity are patterned through and in terms of a distinction between male and female.

Following these insights, we claim that gender acts as a power structure in academia, within the physics field as a space of knowledge production and as workplace, and within the family. As such, it produces and reproduces the career choices of young women.

%%%%%%%%%%%%%
\subsection{The ideal worker and gender boundaries}
One of the examples of the gendered nature of organizations lies in the common gendered perception of the ``ideal worker", based on an image of a man who is free from the burden of taking care of children and family and can put his entire time into work \cite{acker1990}. Acker claims that occupations and hierarchies assume a universal, intangible employee. This employee is, in fact, a man: a man's body, masculine sexuality, control over emotions, and minimal responsibility for reproduction. Images of the male body and masculinity that are dominant in organizational processes marginalize women and contribute to the preservation of gender segregation in organizations.

This claim, which is certainly valid with regard to academic institutions, which were established and took shape in an era when most staff members were men and most of their spouses were housewives. Bagihole and Goode \cite{bagilhole2001} examine Acker's claims in the context of academia. They argue that there is a standard model of academic career. This model is far from being gender-neutral. Instead, it is embedded in a masculine culture and based on a patriarchal support system. Bailyn \cite{bailyn2003} studied the academic careers of senior female faculty in MIT. She argues that the belief, shared by women and men academics alike, is that the only to be a first-rate scientist is to strive to be the ideal, perfect academic, for whom work is the total priority, and for whom there are no outside responsibilities and interests. This may explain why most of these senior women faculty are not married and have no children.

Benschop and Brouns \cite{benschop2003}, who examined gender aspects in academic institutions in the Netherlands, claim that in the basis of the scientific quality standard lies the ``Olympus" model of science, in which the dominant representation of the brilliant researcher is that of a young man at the top of the Olympus, away from the practices of daily life, rooted in the ivory tower of academia. This model is one-dimensional, gender-biased, not open to variance, and may alienate women who do not find themselves in it.

The perception of a profession as male or female is also influenced by the extent to which an occupation allows or does not allow to combine family life with a career. This component was found to be a significant factor in the decision-making of women, who completed PhD in Israel with an excellent grade, whether to pursue an academic career in science \cite{gofen2011}.

Lamont and Molnar \cite{lamont2002} review the idea of ``boundaries" and how it explains various inequalities - class, race, and gender. On the latter, their focus is on how gender and sexual categories shape expectations and work life. In this context, boundaries are defined as ``the complex structures -- physical, social, ideological and psychological -- which establish the differences and commonalities between women and men, shaping and constraining the behavior and attitudes of each gender group" \cite{gerson1985}.  Violation of gender boundaries often leads to punishment and stigmatization in the workplace \cite{epstein2004}.

%%%%%%%%%%%%%
\subsection{Gender inequality}
The attempt to understand the preservation of inequality in academia in general, and in sciences in particular (STEM), has given rise to a growing body of research. Current explanations deal with conscious and unconscious processes taking place at the individual, organizational and social levels.

These explanations can be divided into three categories: Explanations concerning the preservation of the gendered power structure in society, including in organizations, that is also reflected in the field of science, e.g., \cite{acker1990,connell1987}; Explanations concerning organizational structures, processes and practices in academia that are biased toward men, their way of life and merits (nowadays, these explanations are also presented under the organizing concept of ``unconscious bias"; for LERU review, see \cite{gvozdanovic2018}); Explanations concerning organizational culture and climate, which belong, to some extent, to the previous category, but are dealing more with conscious practices and behaviors that have a negative impact on women in academia (chilly climate, sexual harassment and micro-aggressive behaviors toward women) \cite{johnson2018}.

The concept of ``unconscious bias" provides a comprehensive explanation for the preservation of a gendered power structure that favors men over women in various intersections and aspects of academic careers \cite{gvozdanovic2018,atewologun2018}. Knowledge in the field, which has been extensively studied over the past decade, is developing in the area of experimental psychology in research on cognition, perception and social behavior, and in other fields, such as neuroscience and behavioral economics \cite{newell2014}.

Unconscious bias serves as an ``one-size-fits-all" type of explanation for all aspects of inequality in academic career and preference for men over women, starting with recruitment and screening processes (the ``similarity bias" effect, activation of criteria fitting men, such as the number of publications, and recommendation letters in which gender bias against women is present); continue further to working conditions and job characteristics (women receive lower wages than their male colleagues, less research resources , there is a gender bias in funding, in manpower, etc.); and further to unequal practices in the promotion process (promotion processes are biased in favor of men, and preservation of the ``scissors curve," in which the proportion of women decreases as the positions become more senior) \cite{gvozdanovic2018,atewologun2018,kulis2002,long2001,wolfinger2008,gofen2011}.

%%%%%%%%%%%%%
\subsection{Women's career decision making}
Women pursuing non-traditional careers face many obstacles and constraints that can limit or impede their career development. Those who wish to pursue an academic career in masculine fields must often overcome the absence of role models, weak self-efficacy beliefs, uncertain outcome expectations along with cultural and institutional barriers.

Focusing on STEM fields in academy, previous research reveals a high attrition rate for women before and during the postdoctoral studies, a key period towards academic careers, where the numbers of women decrease dramatically \cite{goulden2009,gofen2011,carmi2011}.

Choosing an academic path is a risky decision. The results from a meta-analysis of the research on 16 different types of risk taking indicate that male participants are more likely to take risks than female participants, in almost all types of risks \cite{byrnes1999}. The tendency of women to take fewer risks may explain the lower rates of women who choose the academic path, given its high demands and job insecurity.

Women's career decision to pursue a non-traditional career path, such as physics, could be also explained (following Social Cognitive Career Theory (SCCT) \cite{lent1994,kanny2014}) by their lower self-efficacy (in average).  Low levels of self- efficacy beliefs among women pertaining  to science have been implicated in contributing to the limited number of women earning post-graduate degrees and holding academic appointments \cite{chemers2001,lent1991,grunert2011,falk2017}.

These theories and explanations for decision making are, however, partial and share the same problems as other sociological theories of career decision making (``Trait Theory", ``The Developmental Model" and ``The Social Learning Theory", see \cite{hodkinson1997}). All of them retain a strong sense in which decision-making is fundamentally an individual process, it can contain large elements of technical rationality and it remains within the influence or the control of individuals \cite{hodkinson1997}.

Following Bourdieu and others, we claim that women career decision-making is affected by both their agency and context factors.  The literature on decision making may help to understand women's considerations in choosing an academic career  \cite{nielsen2016}, but consent is never freely or neutrally given in situations of inequality \cite{beddoes2014}.

To better explain the power structures that influence women decision making over an academic career, it is helpful to use Bourdieu's field theory \cite{bourdieu1984,bourdieu1993,bourdieu1990} (see also \cite{hodkinson1997}). According to Bourdieu, agents do not continuously calculate according to explicit rational and economic criteria. Bourdieu uses the agency-structure bridging concept of field. A field can be described as any historical, non-homogeneous social-spatial arena in which people maneuver and struggle in pursuit of desirable resources. In his words, ``it is the state of the relation of force between players that defines the structure of the field" \cite{bourdieu1992}. The `players' within the field are various and they have different resources and power, which make up `the relation of force'. For Bourdieu, each stakeholder brings capital to the field, which can be economic, social, cultural or symbolic.

But, as we learned from decades of sociological studies, those various types of capital are all gendered, and gender operates as an asymmetric capital. Therefore, while masculinity operates in favor of men, femininity does not operate in favor of women \cite{weitz2001}. Moreover, following Bourdieu's theory, the recognition of the limits of what is possible or what is not possible, encapsulated in the decisions of men and women, shapes their aspirations and career paths in a different way. Women's decision making is shaped within a gendered power structure.

%%%%%%%%%%%%%%%%
\section{Methodology}
\subsection{Mixed method research paradigm}
In this study, we used the mixed methods research paradigm, an intellectual and practical synthesis based on qualitative and quantitative research. The mixed methods recognizes the importance of traditional quantitative and qualitative research but also offers a powerful third paradigm choice in order to provide the most informative, complete, balanced, and useful research results \cite{glesne1992,denzin1998,johnson2007}. In parallel, to structure the research tools and analyze the research findings, we used  feminist research approaches and theories that provide framework and tools for looking into women's lives \cite{reinharz1992,devault1999,krumer2014}.

Within our mixed method research, this paper is based mainly on qualitative data collection and analysis, while concurrently recognizing that the addition of quantitative data and approaches into the research contributes to a rigorous understanding of the social phenomena that we study.

\subsection{Qualitative data collection}
The first stage of the research was qualitative. We conducted 25 in-depth interviews with female physics Ph.D. students studying at six research universities in Israel  \cite{footnote:univIL}. Given the small size of the community (there are about 60 female physics Ph.D. students in Israel in a given year), we had to reach out to many of them. The researchers communicated with the students first through an email request. We then used personal connections within the physics community and snowball methodology to achieve a high response rate. No student refused to be interviewed; some even agreed to be interviewed while in maternity leave. The interviews were conducted face to face in the students' offices, labs, and sometimes in their homes or at a coffee shop, according to their request. All interviews were recorded.

Overall we interviewed 25 female physics Ph.D. students. The age of the interviewees ranged from 26 to 36. Most of them were married or in a relationship (21) and only a few were single (4). Twelve were mothers, with 1-4 children, mostly babies or toddlers. Fifteen were experimental physicists and ten were theorists. With regard to their spouse's occupation, 10 were engaged in the fields of computers/engineering/exact sciences in the industrial or private sector, 8 were Ph.D. students in physics, similar to their spouses, and the other had various occupations.

To understand the next phase of the academic career, the postdoctoral path, we also conducted 13 interviews with Israeli female postdoctoral fellows abroad. We reached out to them by our social network within the physics community and by using snowball methodology. Most interviews were conducted online (using Skype), while the scientists were abroad, conducting their research. Most interviewees were in their mid-thirties, all of them were in a relationship, and most of them (11) had children. The number of children ranged between one and four, with two the most common. The majority were admitted to postdoctoral studies in leading institutions in the US and Europe. Since we followed their career in the years following the research we can also note that the post-doctoral duration for most of them has been four years.

The aim of the interviews was to understand the individual and institutional factors impacting their career decisions, whether to pursue an academic career in physics or to leave the academy to a different career path. We listened to their description of their academic path, their choices and their ability to create the future they desire.

All interviews with PhD. students and postdoctoral fellows were recorded, transcribed and thematically analyzed using, at the first stage, ATLAS.ti software and, at the second stage, qualitative research analysis based on the ``Grounded theory" model \cite{denzin1998}.

\subsection{Quantitative data collection}
The second phase of the research included quantitative data collection. Following the interviews with the female students, we wanted to have a broader, representative data regarding all physics graduate students, male and female, so we decided to have a nationwide survey.

The research questionnaire was compiled by the research team in consultation with researchers at the American Institute of Physics (AIP), which has been researching student attitudes toward physics for a decade. The research questionnaire that was formulated is partly based on the tools developed at AIP for research in the field, while adapting it to the Israeli context and to the research questions that interested us (see, for example, \cite{pold2020}). The questionnaire included 106 questions, of which 6 were open ended.  The topics included: students' socio-demographic background, academic study track, attitudes regarding the academic environment, success indicators, combining family and studies, future employment expectations and intentions, desire to have an academic career, considerations in favor of and against postdoctoral studies, and aspects of discrimination and sexual harassment during the academic studies. Some findings of the survey are beyond the scope of this paper and are reported elsewhere \cite{nir2020}.

Physics graduate studies are only possible at few universities in Israel. Therefore, to conduct the research, we reached out to the Israeli Physical Society (IPS) for partnership and support in our study. Through the IPS, we were able to reach all (six) physics faculties in the Israeli universities that have a Ph.D. track in physics: Bar-Ilan University, Ben-Gurion University, Hebrew University, the Technion, Tel Aviv University and Weizmann Institute of Science. Together with the IPS, we approached the six physics deans and asked for their help in distributing the survey. Indeed, all university deans forwarded the survey request to their Ph.D. physics students. The deans also shared with the research team their data about the number of active students by gender. Following the data collection we were able to have the final numbers of physics Ph.D. students in Israel by gender: N=404 students, of whom N=64 women and N=340 men (in 2019).

To enhance the response rate, we promised all students full anonymity, distributed the survey through the faculties mailing lists, and reached out to students to encourage them to answer the questionnaire. We also gave all participants a 15\$ card to buy books as a thank-you gift. We managed to receive a very high response rate: 66\% of the overall population of students in the country (267/404), with an even higher response rate -- 94\% --  for women (60/64), and 61\% response rate for men (207/340). We received answers from students in all six universities. The population size and response rate by institution are presented in the Table \ref{tab:phd}.

\begin{table}{b}
 \begin{center}
   \begin{tabular}{cccc}
   \hline
   Institution & Students & Respondents & Response rate \\ \hline\hline
  Weizmann Institute & 109 & 71 & 65\% \\
  Bar-Ilan University & 67 & 50 & 74\% \\
  Hebrew University & 65 & 40 & 62\% \\
  The Technion & 65 & 39 & 60\% \\
  Tel Aviv University & 64 & 39 & 60\% \\
  Ben-Gurion University & 34 & 28 & 82\% \\
  {\bf Total} & {\bf 404} & {\bf 267} & {\bf 66\%}
   \\ \hline
  \end{tabular}
  \caption{Physics Ph.D. students in Israel: the overall population and the number of respondents by institution.}
\label{tab:phd}
 \end{center}
\end{table}

The maximum error for the entire population is 3.6\%, among women 3.2\% and among men 4.3\%. Due to the representation of women in the sample, the total data of the students was weighted by gender. Data analysis was performed using variance to proportions analyzes, given small populations.

%%%%%%%%%%%%%%%%
\section{Findings}
%%%%%%%%%%%%%
\subsection{The postdoc as a ``deal"}
Exploring the student's expectations of an academic career illustrates their image of the field. We asked all Ph.D. students that state they would pursue an academic career, why they would choose this career path.
Their answers were somewhat surprising. We found that love for physics and a deep interest in this field were the most common answers (inter alia, loves research, loves exploring, loves basic research, this is my dream, it fits my character). The next common answer was the academic freedom (the ability to conduct my own research, independence, no bosses or customers). The third common answer was related to the work conditions (favorable conditions, tenure, job security, prestige, social status and leadership capacity).

Based on both the survey and the interviews, we thus find that the main benefits of the academic career ``deal" are the ability to engage in scientific research in a fascinating field, intellectual freedom to explore and be creative, independence in choosing what and how to do research, freedom from bosses or clients, and working conditions that guarantee employment stability and (reasonable) economic well-being. The deal does not include quick enrichment, but it includes the prospect of groundbreaking scientific discoveries (and worldwide fame alongside them), as well as prestige that comes with being part of the exclusive club of the intellectual elite.

In light of these career benefits, we found that, at the crossroads of pursuing a postdoc, the academic career is considered as a ``deal," which has three main components: personal-marital, professional-occupational, and financial. Young women are realistically examining the components of this ``deal": what it offers them and what prices they will have to pay. In accordance with these considerations, the decision is made. This does not imply that the considerations are all ``rational" or ``objective." Undoubtedly, the decision involves feelings and thoughts, realistic and unrealistic expectations, perceptions of academic institutions and labor market, and aspiration of professional and personal future, but the bottom line is that all of the above are merged into one informed decision, whether to go for a postdoc abroad as a necessary step for an academic position, or to quit the academic race at the current stage.

Embarking on a postdoctoral career is a significant, even dramatic step in the lives of young women (and men), and requires a will to make significant changes in many aspects of life for a long period of time. In order to obtain a tenure-track position in physics, two postdoctoral periods are often required, {\it i.e.} a cumulative period of about four years abroad. Therefore it has personal, family, professional and financial implications for both women and men, and occurs under conditions of uncertainty and job insecurity. Women explore all of these aspects while examining the option of a postdoc.

On the personal and family level: relocation to a foreign country is required, which includes in many cases the need to integrate children into new schools and kindergartens while learning a new language, and integrating into a new social and cultural environment. On the professional level: first, the candidate must find an academic mentor and an institution willing to host her for the postdoctoral period. This task requires talent, self-marketing skills, willingness to travel abroad, and an effort to become acquainted with suitable scientists and institutions. During the postdoctoral period, the candidate is required to prove herself again, publish, make a good impression on the relevant professional community and prove her capability as an independent scientist. After all these efforts, the postdoctoral researcher is not guaranteed to get an academic job, as competition for jobs is high, and the chances are unknown. On the financial level, there is the concern for making a living. The scholarship during the postdoctoral period is significantly lower than physicists' average wage in the labor market. It does not include social benefits and accrual of future rights, such as pension and education fund. Most scholarships are designed to allow a single person to live a modest life, and usually do not suffice for a family. The costs of living abroad for a family with young children may be significantly higher than the average scholarship. It means they have to fund postdoctoral studies via savings or via family support, in a time of their lives when they are expected to be financially independent. Moreover, relocation abroad may impair the spouse's income, employment continuum and skills. For some professions, it is hard to find a parallel job abroad (for example, lawyers or military officers); in some cases, immigration-related restrictions do not allow the spouse to work.

All of these components are considered by the women and their spouses when making the decision whether to go on postdoctoral studies.

%%%%%%%%%%%%%%%%%%%%%
\subsection{The gendered aspects of the deal}
While both women and men examine the ``deal" terms at a similar stage of their lives, it is clear that among women, the gendered power structure creates different expectations and extra hurdles, which make their decision to go for a postdoc more challenging.

Gendered power structure influences women's academic careers in physics in numerous ways. First, women face unequal competition in physics as a masculine field. Second, couples prioritize man's career over the woman's career. Third, postdoctoral career path is socially perceived as a disruption of the gender order. Women justify this non-normative path by demanding of themselves exceptionally high standards of academic excellence. We claim that these standards of excellence operate as a hidden component within the gender regime that justifies women's decision to go for postdoc.

%%%%%%%%
\subsubsection{Unequal competition in physics as a masculine field}
Physics is a masculine field, characterized by supremacy of a white male majority, with masculine culture and masculine public image of the field. Women's integration into this field is relatively new and the gender imbalance in the field is significant worldwide \cite{cochran2019}.  Following Bourdieu's theory \cite{bordieu1977,bordieu1986}, we study physics as a social field. Social field is a patterned set of practices within a broader social space, which suggests competent actions in conformity with rules and roles. It is a playground or battlefield in which actors, endowed with a certain field-relevant capital, try to advance their position.

Traweek \cite{traweek1988} conducted an in-depth anthropological study of the world of physics, tracing the culture of high energy physicists in one of the leading laboratories in the US. Traweek describes physics as a male domain and claims that masculinity is an organizing principle of the physics laboratory, as part of the common perception that science is a field of ``individual great men." The field of physics is highly competitive. To be accepted as a member of the physics community, you have to be very committed, charismatic, highly motivated, dominant and aggressive. Only the strongest and brightest manage to overcome these obstacles or, in her words, ``Only the blunt bright bastards make it".
Thirty years after Traweek's work, the field of physics is still masculine and highly competitive. The gendered labor market in the physics field is clearly reflected in both our survey findings and the interviews.
Along the last decade, females constitute only 16-17\% of the B.Sc., M.Sc. and Ph.D. physics students in Israel and there is no sign for positive change. At the faculty level nationwide, females constitute only 6\% of the overall staff in all physics faculties \cite{jona2019}.

The marginal position of women in the field of physics is evident not only quantitatively, but also qualitatively, in the women's experiences. Most female Ph.D. students (63\%) reported having experienced gender based discrimination during their studies (compared to 7\% of men, $p<0.01$). Moreover, one in every five women (21\%) reported being sexually harassed during their studies (compared to 2\% among men, $p<0.01$). Of these, half were harassed more than once. Only a minority of victims reported it to competent entities.

In the male-centric culture, the fact that women are different plays against them. The disadvantage becomes more prominent when they become parents. While women are expected to become the main caregiver, their men colleagues are expected to follow their career as they had done before they became parents. Timing works to the detriment of women since, while they need flexibility in order to raise young children, they have to prove themselves to a highly demanding system that does not stop for a moment \cite{ceci2010}, a system in which there is no such thing as a ``good" time to have children.

Although many of the women said that they strive to implement an egalitarian model of childcare at home, most of them also report they spent longer hours in childcare and child-related work compared to their partners, and that this comes at the expense of their studies.

Analyzing the survey data, we found that, despite of a prominent presence of an egalitarian ideology among physics student families, expressed by women's and men's desire for an equal distribution of roles, it is evident that women carry a greater burden of family work. First, they take a longer maternity leave, a period of time that impedes their studies (69\% of women take a maternity leave of 4 months or more, compared to very short parental leave used by the male students). Women are also going through the pre-birth period, during which many women need to undergo various examinations, and sometimes require medical care and observation, which take time and require a lot of attention (a reality that may repeat itself after the maternity leave as well).

Second, women, more than their male colleagues, reported that due to parenthood, they adopted a more flexible work schedule (60\% of women, 48\% of men, $p<0.05$), and that they learned to be more productive in their studies (40\% of women, 27\% of men, $p<0.05$).

Third, examination of the role distribution structure in these families indicates that, although the fathers are engaged in childcare, there is still gender inequality at home. 100\% of women reported that they are responsible for taking care of their children's needs (57\% are the primary caretaker, while the remaining 43\% share childcare with their spouse equally). In contrast, 28\% of men reported that most childcare responsibilities are imposed on their spouses, 67\% share this responsibility with their spouse, and only a minority of 5\% reported to bear most of the responsibility for childcare.

Moreover, although most women and men reported that they share household chores (67\% and 64\%, respectively), no women is free from this burden (the remaining 33\% reported that the household burden mostly lies on their shoulders), while men either share the burden with their spouses or it is mainly imposed on their spouses (24\% of men stated that this burden is mainly imposed on the spouse).

These findings are not surprising. They are manifest in studies about time spent on housework in western democracies for decades. According to the American time use survey, women spent an average of 2 hours and 15 minutes a day on housework, while men spent 1 hour and 25 minutes \cite{atus2016}.  Women continue to take the primary responsibility for home and family even in the most gender-equal countries \cite{seierstad2015}.

This gendered role division at home is reflected in the academy, by a more significant presence of men at physics labs and offices, in terms of time allocation. Most male physicists (even if they truly believe in gender equality) do not practice parenthood in the same way as women. Though most Ph.D. male students declare that parenthood affected their studies, parenthood is not as significant a variable in their lives as employee, as they follow the social expectation that the family will be pushed aside due to their career demands. At the same time, the usual expectation from women is to do both, to be both dedicated mothers and career women.

As found in research conducted among Israeli fathers, even though there has been a change in recent decades in fathers' involvement in household chores and childcare practices, the parental responsibility for the private sphere is still unequal. This sphere remains feminine. Men spend longer hours at work, and most of the household chores and childcare become their wives' burden \cite{anabi2019}. These findings are in line with studies conducted in other Western countries (see, {\it e.g.}, \cite{doucet2015}) and with the research on women working in male-centric domains, that experience daily battles as competing desires to both be a ``caretaker" at home and develop a professional career \cite{case2013}.

In this reality of unequal distribution of roles at home, parenthood is still, as Ceci and Williams claim \cite{ceci2010}, a significant barrier to women's integration into science professions in academic institutions. In the Israeli context, this concept is embedded in the perception of the family as a central institution in the individual life and part of the national strength, and the perception of a woman as, first and foremost, a mother and wife, and only then as a provider \cite{fogiel1999}.

The interviews show that women's greater commitment to family makes it harder for them to succeed in their studies within a male dominated culture. The young females report being discriminated based on the normative assumption that mothers are less competent and committed than other types of workers, as was documented in previous research addressing the ``motherhood penalty" phenomenon \cite{benard2010}.

T., a mother of a two-year-old girl who was pregnant at the time of the interview, told us about the difficulty of combining studies with motherhood. T. aspires to an academic career but, at the same time, she gives high priority to her family. She tries to live up to the social expectation of her ``doing both." She finds out that this situation puts her at a disadvantage in her daily competition with her male counterparts. Her commitment to family, or what she calls ``this problem," is her problem, and not a problem of her colleagues, all of them male and free from having to live up to the expectation of being the main caregivers for their children. T describes it in a tone of acceptance, but also criticizes it:

``My family, my husband, my marriage and my children are very important for me, they are very high in my priorities... I feel the gender differences (compared to male colleagues), mainly since I have much less time to work than my friends from the lab, and it becomes a big gap... it's like you are competing against those to whom you compare yourself, all the time... It's hard to combine motherhood with anything that is career related, not only in academia, but the competition is a competition with men who don't have this problem."

As a young mother, the difficulty of combining career and motherhood requirements become clear to T. Taking care of her two-year-old daughter takes precious time and harms her ability to successfully compete with other students, her lab colleagues, whose time is at their disposal and are less challenged by family demands.

The women's experience is structured within a gendered labor market, where men and women have to live up to the same expectations at work, in the public sphere, but different expectations at the private sphere. Women's understanding that physics as a field is gendered, that the competition, in which they are and will be required to compete, is unequal, and that their challenge is to be both a mother and a career woman, leads some of them to quit the academic career race. Those who stay in the race understand that they must succeed in these gendered conditions, while competing from an unequal starting point with those who are free from the ``second shift" at home.

%%%%%%%%%%%%%%%%%%%%%
\subsubsection{Prioritizing the man's career}
In Israel, in comparison to other Western democracies, students start academic studies relatively late and marry at a relatively young age. Therefore, Ph.D. students' family situation in Israel has unique characteristics.

Based on the Ph.D. student survey, male and female Ph.D. students in physics are in their early thirties. The average age is 29.7 among women and 31.8 among men. Most women (63\%) and men (70\%) are already married or in a relationship, and more than a third of them already have children (41\% of men and 34\% of women). Moreover, about 25\% of male and female Ph.D. students have two children or more.

This picture is clearly reflected in the interviews. Most of the Ph.D. students we interviewed (21 out of 25) already had spouses, half of them (12) were already mothers of young children at the time of the interview, while the single women declared their desire to become mothers in the coming years.

The majority of spouses of the women we interviewed were working in STEM fields (for example, engineers and computer scientists) or were graduate students in STEM. Most of the female students described their spouses as mostly supportive in their career aspirations and decisions, wanting and willing to help them succeed. Many students described their spouse as one of the factors for their own success (88\% of women and 80\% of men have noted this in the survey, $p<0.01$). However, the interviews clearly indicate that, once there is a spouse, career and family considerations become intertwined and therefore more complex. Career becomes ``spousal" in the sense that a decision made with regard to the man's career affects the woman's career, and vice versa. The couple is considering the impact of their choices on the entire family.

The interviews with both Ph.D. and postdoc female students prominently showed that women give a significant weight to the implications of going for a postdoc on their spouse's career. Women are preoccupied with the questions: Will my spouse be able to find a job or a postdoc abroad (in view of language and visa related limitations)? Will he be able to find work abroad that fits his skills (depending on his professional characteristics, job availability, his ability to adapt to a different job market, etc.)?

A clear example of a ``spousal" career and interrelated career considerations emerges from the story of S., married with two toddlers. S. found it difficult to separate her career aspirations from her partner's aspirations, which she presented as interdependent. The interviewer tried to refine the differences and understand what she wants:

Q: "I'm trying to understand how you see your career, if you didn't have any limitations, where would you like to see yourself?"

A: "Now, it seems that my doctorate is going to be more or less successful and I'll have good results, so yes, I would like to continue in academia and do a postdoc, and I know that there is some institution in Europe... they are looking for people (in our field) and we can both get work there, some sort of a postdoc, so that could be nice. But now it comes to my husband and if he finds a place that he will really love, and there won't be a place for two people there, I'll go look for a job in high-tech or something else, and that would also be perfectly fine, and if I won't find a job in high-tech, I will be a teacher and it will also be fun... It won't be as interesting as research, but it's a job."

S. subordinates her desires to those of her spouse, his career is being given a clear priority within the spousal relationship, although her Ph.D. was good and she recently won an excellence award.

In another case, the counterweight to postdoc was the spouse's desire to stay in Israel. B.G. is freshly married, and pregnant with her first child. She told us that her spouse supports her, but at the same time he does not want to leave Israel and is very connected to his country and his family. B.G. eventually decided to leave academia. It is impossible to determine if it was her spouse who affected the decision against postdoc, but it is clear that his will had a significant weight.

Reinforcement of the findings that emerged from the interviews is found in the survey. While both men and women reported that their spouse's employment considerations play a key role in their decision whether to go for a postdoc, women have given more weight to this issue. 81\% of the women indicated their spouse's ability to find a job abroad as a key consideration, compared to 70\% of the male Ph.D. students ($p<0.05$), and 77\% of women noted the difficulties involved in relocating abroad with the spouse and family as a key consideration, compared to 66\% of men ($p<0.05$).

Thus, it seems that even if the male spouse is supportive and willing to follow the woman abroad, women give great weight to his career, desires and preferences. This reflects, inter alia, the gendered power structure that exists within society as well as in the job market.

The priority and precedence given to the man's career in many families could be based on the understanding of their better chances for higher wage and promotion (in average) at the job market. It was surprising to find that, at this early stage, when most students live on a modest subsistence scholarship, there are already financial gaps in income, in favor of men. The majority of female Ph.D. students (67\%) indicated that their spouse's income from school or work is higher that their income, compared with a minority (28\%) of male Ph.D. students. The male Ph.D. students' wages are higher than their spouse's wages (46\%) or equal to them (20\%, the remaining 6\% have no income).

This power structure in the job market also emerges in interviews that reflect the patriarchal structure of society and power relations within the family. When a man follows his wife abroad, it disrupts the gender order. The priority given within the family to the male spouse's career restricts women from embarking on a postdoc and limits their choice of academic career.

This gendered power structure is so profound that it even affects single women, who do not have a spouse and children. Three Israeli single Ph.D. Students whom we interviewed claim they refrain from embarking on postdoctoral studies, inter alia, due to the concern that they may impair their chances to get married in the future or create potential limitations on a future (spouse's) career.

%%%%%%%%%%%%%%%%%%%%%
\subsubsection{Postdoctoral career path as a disruption of the gender order}
In the Israeli context, there is a requirement to undergo for postdoctoral studies abroad, {\it i.e.}, the student must leave the country for a prolonged period of professional research and development. (The requirement is not formal, but the probability of getting an academic position after a postdoc in Israel is much lower.) Although the common view in contemporary educated circles is that the job market is open and equal for women, the interviews show a much more conservative view. It is evident that the social and family environment perceives postdoctoral studies as an ambitious and non-normative path for women. It is common for women to follow their spouses for a period of work or studies abroad, but the opposite model is still considered non-normative and is perceived as ``feminist" and challenging the common social order, in which the male career is the lead.

Ts. is about to embark on a prestigious postdoc in the US with her spouse. She describes the postdoctoral path as non-normative for a woman, which is why it has to be negotiated with her spouse and be justified against the family system:

``I think a postdoc abroad takes a heavy toll. Usually the husband is older and has a job, and does not want to leave. The easy cases are when the husband also goes for a postdoc, or can work abroad and, in such a case, he wants to leave. Even if the husband is supportive, the broader family wrinkle up their noses and put pressure on me (not to go on a postdoc abroad). If it were possible to do a postdoc in Israel, it would be much easier for women."

M., single, PhD student, believes that women are less likely to go for a postdoc because it is a deviation from the conventional structure of gendered power relations. In the accepted social order, man's career is the significant one, and not the other way around:

``I think that the cultural perception, at least in Israel, is that the woman will follow the man, {\it i.e.}, if the man has to relocate due to work or studies, it is perceived as more natural for the woman to follow him."

M. says the postdoc issue has come up in her previous relationship, and although her former spouse's attitude regarding this issue was positive, his social environment was against this move and regarded it as non-normative. The idea of him, a man, relocating abroad for the benefit of his wife's career, was met with criticism and astonishment by his colleagues.

%%%%%%%%%%%%%%%%%%%%%
\subsubsection{Self-expectations for excellence: a hidden component in the gender power structure}
As found in previous studies, women must demonstrate stronger abilities than men in order to be recognized as equally good. In male dominated disciplines, for women to be considered good, worthy of employment and promotion, they must be better than their male peers \cite{heilman2008,kaatz2014}.

In a gendered workforce, when women compete over an academic career in a masculine field, having to overcome gender discrimination and motherhood penalty, prioritizing their husband's career at home, and taking a greater share of the childcare, what justifies their decision to go on a postdoc abroad, or to embrace the academic ``deal"?

Based on our research findings, we claim that, to justify this deviation from the ``gender order", women are pushing themselves to excel. Excellence was brought out as a justification that allows deviation from the norm and disruption of the gender order. Many women stated that it is considered obvious when a man goes for a postdoc and his wife goes with him, even if it requires that she gives up her career. In contrast, for a woman to embark on a postdoc with her partner joining her, special conditions must be met. One of the unspoken conditions, mentioned repeatedly along the interviews, is being an excellent student. This is how B.G., married and pregnant, explains why she decided to leave academia for a job in the industry after completing her Ph.D.:

``Women are also affected by their partner, not that men are not, but to a certain extent, when a man thinks about going for a postdoc, his spouse is excited to follow him... it's an adventure, (on the contrary) a woman waits to hear the man's opinion, and if he says no, then there should be a really good reason, for example, when (your) doctorate is brilliant and the supervisor wants you to travel for postdoc... then, maybe then."

Thus, when a man relocates abroad following his wife's career, it is viewed by the interviewees as a disruption of the gender order. To deviate from the norm, the woman must be excellent. It is not enough for her to be a good or even a very good student, and she probably cannot afford a postdoc if she is an average student.

The excellence must be reflected in numerous aspects that are interconnected: her academic achievements; her supervisor's evaluation of her; the professional group perception of her academic potential; and her ability to receive postdoc offer from a top institute.

Excellence has two functions, in the public sphere and in the private sphere. In the public sphere, excellence allows the students to feel worthy to face professional competition against their male colleagues. In the private sphere, it justifies (to the spouses and to the women themselves) their choice of a non-normative academic career path for a woman, which includes going abroad for a postdoc and subordinating the family in accordance with their career needs.

A very clear example of excellence as a justification for disruption of the gender order comes out from the story of A., recently married with a baby. A. said she had already decided when she started her Ph.D. studies that she would not pursue a postdoc in a place that is ``not good". She says that, in the beginning of the relationship, her husband and she agreed that she would pursue postdoctoral studies if she is accepted into a prestigious institution.  In return, he promised that he would be willing to leave his job to support her career. This is her answer to the question of whether she would like to embark on post-doctoral studies:

``Yes, very much! But it is contingent upon me getting a good postdoc! I mean, not a postdoc from the University of Nowhere, I don't know, something like that, it should be a good postdoc! Because basically, my husband will come with me and will have to take a leave without pay, which is also not so trivial at work... I won't drag the entire family if it's a postdoc that will get me nowhere, you know, it should be a good and lucrative postdoc, so I would have some motivation to return to Israel. I really want to be in academia, and I think that science is just the best thing there is, in my opinion, and it's something I want to do all my life."

Another example rises from the story of V., married and mother of three, who is about to graduate. She shares her doubts about the future. She claims her spouse supports her postdoc aspirations, however, if they relocate, he would have to give up a job that is a significant part of his life. Therefore, for V., going for a postdoc takes a high price for her partner's career. After some deliberation, V. decided to do her first postdoc in Israel. Only if it is successful, she will go abroad for another period.

Academic career is viewed as a highly demanding and competitive path, but also as a masculine, non-normative path for a woman, by the social and family environment. Women in physics understand that they must compete under unequal conditions of a gendered labor market, and that they must excel in order to be perceived as equal. Under these conditions, excellence operates as a hidden component, within the unseen gender regime, that justifies and allows deviation from the gender order by going for a postdoc abroad and prioritizing women's careers over their husbands'.

Whereas for married men, motivation is all they need to go for a postdoc abroad, for women motivation is not enough. Married women need a very good justification for their non-normative career choice and for their husband to follow them. This might be one of the dominant reasons for the low number of women pursuing an academic career in physics and the gender imbalance in the academic staff in physics.

%%%%%%%%%%%%%%%
\section{Summary and discussion}
In this study, we look at career decision-making in physics as a multilayered and multi-dimensional phenomenon. Studying both context factors and agency, we find that the academic career in physics offers a ``deal," which has three main components: personal-marital, professional-occupational, and financial. Young women are realistically examining the terms of this ``deal," what it offers them and what prices they will have to pay. The decision is made in accordance with these considerations but, contrary to men, women are operating within a gender power structure that navigates their decisions in a different way.

While both women and men consider the ``deal" terms in a similar stage of their lives, among the women, the gendered power structure creates different expectations and extra hurdles, which make the decision to pursue an academic career and to go for a postdoc more challenging. Our findings reveal the multiple and hidden ways in which gender operates as a power structure in the labor market within the physics field, in the spousal relationship within the private sphere, and in the social norms and expectations within society, putting up a barrier to women's academic careers. This latent power structure influences women's decision-making and experiences in several ways. In the academic field, it produces unequal competition in a male-dominated playground, where women struggle to succeed as physicists and as mothers, but are viewed as less devoted workers because of their parental commitment. In the social and family spheres, choosing a demanding academic career is seen as a non-normative trajectory for women and as disrupting the gender order. Within the private sphere, women carry a greater share of the childcare and family work and, moreover, give priority and precedence to their partner's career and preferences.

Women justify this non-normative path by raising their self-expectations for excellence. They feel that they must excel in their research and exhibit exceptional achievements. We claim that excellence operates as a hidden mechanism within the gender regime, that can justify women's decision to go for postdoc, but can also operate as an exclusionary mechanism that prevents many talented young women from choosing an academic career in physics. We should stop thinking about women as giving up the academic career ("the leaky pipeline" discussion), but rather as choosing career paths that align better with their expectations as women and workers.

If we want the academy to be more gender-balanced, so that women are no longer a token minority, we should tackle the many obstacles women face, both in physics as a field and within the family circle. We should make physics more appealing for women, knowing that their choices are made within a gendered structure and that the academic path is more demanding for them. If academia does not act in this direction, it will lose talented females to the global tech companies, which in recent years have made intensive efforts to change the gender balance among their employees and integrate more women into diverse work teams.

Further research is needed to support these findings and to explore hidden barriers to integrating women into academic careers, in physics and other scientific disciplines where they still constitute a token minority.

%%%%%%%%%%%%%%%%
%%%%%%%%%%%%%%%%
\subsection*{Acknowledgements}
We thank Sharon Diamant-Pick for her help in conducting the survey and in analyzing the interviews.
YN is the Amos de-Shalit chair of theoretical physics. This research is supported by grants from the Israeli Ministry of Science and Technology, from the Estate of Rene Lustig and from the Estate of Jacquelin Eckhous.
\vspace{6 pt}

%%%%%%%%%%%%%%%%%%%%%%%%
%%%%%%%%%%%%%%%%%%%%%%%%

\end{document}